\documentclass[prl,twocolumn,showpacs,amsmath,amssymb,floatfix]{revtex4}
\usepackage{amsfonts,mathtools,float}
\usepackage[colorlinks=true,citecolor=blue]{hyperref}
\usepackage{mathpazo,times}

\def\doibase{http://dx.doi.org/}
\def\Real{\mathbb{R}}
\def\e{\mathrm{e}}
\def\d{\mathrm{d}}
\renewcommand{\H}[0]{\mathcal{H}}
\def\sign{\mathop{\rm sign}\nolimits}

\def\re{\mathrm{re}}
\def\im{\mathrm{im}}
\renewcommand{\Re}[0]{\mathop{\rm Re}\nolimits}
\renewcommand{\Im}[0]{\mathop{\rm Im}\nolimits}
\def\tr{\mathop{\rm tr}\nolimits}
\renewcommand{\H}[0]{\mathcal{H}}
\def\pvint{\mathop{\int\kern-1.02em-\kern0.2em}\limits}

\def\bse{\begin{subequations}}
\def\ese{\end{subequations}}

\def\inst#1{$^{#1}$}

\begin{document}
\title{Resonant optical pulses on a continuous wave background in two-level active media}
\author{Sitai Li\inst{1}, Gino Biondini\inst{1,2}, Gregor Kova\v{c}i\v{c}\inst{3} and Ildar Gabitov\inst{4}}
\affiliation{%
\inst{1} Department of Mathematics, State University of New York at Buffalo, Buffalo, New York 14260, USA\\
\inst{2} Department of Physics, State University of New York at Buffalo, Buffalo, New York 14260, USA\\
\inst{3} Department of Mathematical Sciences, Rensselaer Polytechnic Institute, Troy, New York 12180, USA\\
\inst{4} Department of Mathematics, University of Arizona, Tucson, Arizona 85721, USA
}

\begin{abstract}
We present exact $N$-soliton optical pulses riding on a continuous-wave (c.w.) beam that propagate through and interact with a two-level active optical medium.  
Their representation is derived via an appropriate generalization of the inverse scattering transform for the corresponding Maxwell-Bloch equations.  
We describe the single-soliton solutions in detail and classify them into several distinct families.  
In addition to the analogues of traveling-wave soliton pulses that arise in the absence of a c.w.\ beam, 
we obtain breather-like structures, 
periodic pulse-trains and rogue-wave-type (i.e., rational) pulses, whose existence is directly due to the presence of the c.w.\ beam.  
These soliton solutions are the analogues for Maxwell-Bloch systems of the four classical solution types of the focusing nonlinear Schr\"odinger equation with non-zero background, 
although the physical behavior of the corresponding solutions is quite different.
\end{abstract}

\maketitle

\section{I.~ Introduction}
%

The study of resonant interactions between coherent light and two-level media has been an active area of research for the last forty years.
Typically, these interactions are described by Maxwell-Bloch systems of equations (MBEs),
which are completely integrable in certain limits, 
which means that many analytical tools can be brought to bear on these problems.
Indeed, the initial-value problem (IVP) for non-degenerate two-level systems was solved in \cite{lamb1973,akn1974},
using the inverse scattering transform (IST) \cite{as1981,nmpz1984}
and used to characterize the phenomenon of self-induced transparency \cite{mccallhahn1,mccallhahn2,Lamb1,Lamb2,physrep1990}.
The theory was also later generalized to MBEs in the so-called lambda configuration in \cite{lambda1,lambda2,lambda3,lambda4,tmp1985,bgk2003,CPA2014},
and was also used to provide a description of polarization switching phenomena \cite{bgk2003,gk1,gk2}.
The soliton-radiation interactions were also studied using reduced MBEs via a combination of IST and Darboux transformation~\cite{msr1988,rt1993}.

On the other hand, most studies in the literature consider the case of localized optical pulses,
i.e., optical fields which vanish in the far past and future.
The goal of this work is to study optical pulses in MBEs with nonzero background (NZBG), 
i.e., with the optical field limiting to a finite, nonzero value as $t\to\pm\infty$.
Understanding the behavior of such systems is key to obtain a mathematical description of
slow light phenomena \cite{hhdb,ldbh2001}.
(To avoid confusion, we point out that in some earlier works the term ``non-zero background'' was used to indicate that 
some entries of the density matrix do not vanish as $t\to\pm\infty$.
Note, however, that the density matrix can never vanish as a whole.
Therefore, here we prefer to reserve the term NZBG for cases in which the optical field is also non-vanishing as $t\to\pm\infty$,
denoting cases in which the optical field vanishes 
as $t\to\pm\infty$ as zero background, or ZBG.)

Some explicit solutions of two-level MBEs, 
coupled MBE systems (the so-called ``lambda'' configuration) and related systems 
(e.g., mixed systems of coupled Maxwell-Bloch and nonlinear-Schr\"odinger equations, or Maxwell-Bloch and Hirota equations) 
with NZBG have recently been produced using Darboux transformations \cite{pb2001,RVB20051,RVB20052,xphc2016,hxpcd2016,hxp20121,hxp20122,lhp2013,wlqz2015}, 
but no comprehensive theory is available.
Moreover, previous studies on two-level MBEs with NZBG~\cite{pb2001,xphc2016,hxpcd2016} only considered the so-called sharp-line limit 
(in which the detuning function is taken to be much narrower than the spectral width of the pulse), 
and are also related to complicated (and possibly unphysical) material preparations.
Here, we use the IST for non-degenerate two-level systems with NZBG 
and \textit{a general detuning function}
to derive and discuss several new exact solutions
which display novel kind of behavior.
We also discuss how the choice of the NZBG affects the solutions,
and we show how the solutions of the present work differ from those in Refs.~\cite{pb2001,xphc2016,hxpcd2016}.

Specifically, we consider non-degenerate two-level optical systems with one excited state and one ground state.
The governing equations are the scalar MBEs,
which in the light-cone reference frame can be written in matrix form as \cite{tmp1985}
\vspace{-0.4ex}
\bse
\label{e:MBE}
\begin{gather}
\rho_t= \big[i\lambda \sigma_3 + Q,\rho\big]\,,
\\
Q_z = - \frac{1}{2}\int\nolimits_{-\infty}^\infty\big[\sigma_3,\rho\big]\,g(\lambda)\d\lambda\,,
\label{e:MB2}
\end{gather}
\ese
where 
subscripts $t$ and $z$ denote partial derivatives,
$z= z_\mathrm{lab}$ and $t= t_\mathrm{lab} - z_\mathrm{lab}/c$ 
are the propagation distance and the retarded time, respectively, 
$c$ is the speed of light in vacuum,
$[A,B] = AB - BA$ is the matrix commutator,
$\sigma_3$ is the third Pauli matrix,
and $g(\lambda)$ is the detuning function due to inhomogeneous broadening
(e.g., as due to Doppler effect), with 
$\lambda$ being the frequency detuning parameter.
The Hermitian $2\times2$ density matrix $\rho(t,z,\lambda)$ describes the quantum state of the medium,
In particular, the $(1,1)$ and $(2,2)$ entries of $\rho$ denote respectively the population of atoms in the excited and ground state.
Since \eqref{e:MBE} is invariant under the transformation $\rho(t,z,\lambda)\mapsto\rho(t,z,\lambda)-\rho_o(z,\lambda)\,I_2$,
where $I_2$ is the $2\times2$ identity matrix and $\rho_o(z,\lambda)$ is an arbitrary scalar, 
without loss of generality we can take $\tr\rho =0$ for all $z\ge0$ and for all $\lambda\in\Real$.
Choosing normalizations so that $\det\rho = -1$ as usual,
the matrices $Q(t,z)$ and $\rho(t,z,\lambda)$ in \eqref{e:MBE} are then
\vspace*{-0.6ex}
\begin{equation}
\label{e:Qrho}
Q = \begin{pmatrix} 0 & q \\ - q^* & 0 \end{pmatrix}\,,\qquad
\rho = \begin{pmatrix} D & P \\ P^* & -D \end{pmatrix}\,,
\end{equation}
%
where $q(t,z)$ is the optical field amplitude corresponding to the transitions between the two quantum states, 
the asterisk denotes complex conjugate,
$D(t,z,\lambda)$ relates to the ratio of populations of atoms in the two states,
and $P(t,z,\lambda)$ is the polarizability of the medium.
Even though the IST can be carried out for arbitrary $g(\lambda)$,
for concreteness all the solutions presented below are obtained by taking the usual choice of a Lorentzian distribution \cite{akn1974},
namely,
\vspace{-0ex}
\begin{equation}
\label{e:Lorentzian}
g(\lambda) = \epsilon\big/[\pi(\epsilon^2 + \lambda^2)]\,,\qquad 
\epsilon >0.
\end{equation}
(Other choices are possible, of course.)  
A special case is obtained when there is no detuning, 
and all atoms are resonant at exactly the same frequency.
This case is usually called the sharp-line limit, and corresponds to the limit of \eqref{e:Lorentzian} as $\epsilon\to0$, 
in which case 
$g(\lambda)$ yields the Dirac delta distribution $\delta(\lambda)$.

The main result of this work is a rich family of soliton solutions, 
and their characterization in terms of the unfolding of the discrete eigenvalue in the spectral plane.
All of these solutions are novel to the best of our knowledge.
The outline of this work is the following.
In section~II we present the essential elements of the IST for the MBEs~\eqref{e:MBE} with NZBG.
In section~III we give the general expression for the soliton solutions.
In section~IV we discuss in detail the various classes of one-soliton solutions.
Finally, section~V concludes this work with a few final remarks.
Further details about the IST and the soliton solution formulae are given in Appendix, 
together with detailed discussion on alternative choices of boundary conditions.

\kern\smallskipamount
\section{II.~ IST for MBEs with NZBG}
%

The MBEs~\eqref{e:MBE} are the compatibility condition of the Lax pair~\cite{akn1974}
\vspace{-0.4ex}
\bse
\label{e:LP}
\begin{gather}
\phi_t = (i\lambda \sigma_3 + Q)\,\phi\,,
\label{e:LP1}
\\
\phi_z = \displaystyle\frac{i\pi}{2}\H_\lambda\big[\rho(t,z,\lambda')g(\lambda')\big]\,\phi\,,
\end{gather}
\ese
where $\lambda$ plays the role of a spectral variable,
$\phi(t,z,\lambda)$ is a matrix eigenfunction and 
$\H_\lambda[f(\lambda')]$ is the Hilbert transform
given by the Cauchy principal value integral
\vspace{-0.5ex}
\begin{equation}
\H_\lambda[f(\lambda')]= \frac1\pi \pvint \frac{f(\lambda')}{\lambda'-\lambda}\d\lambda'\,.
\label{e:Hilbert}
\end{equation}
(The prime in the argument of the Hilbert transform denotes the integration variable, \textit{not} differentiation.)
The Zakharov-Shabat scattering problem~\eqref{e:LP1} is the same as that for the focusing 
nonlinear Schr\"odinger (NLS) equation~\cite{ZS1972}. 
Thus, the direct and inverse scattering for Eqs.~\eqref{e:MBE} coincide with those for the focusing NLS
equation, and the only difference in the IST is the spatial dependence of the spectral data.

The IST for the focusing NLS equation with ZBG and NZBG was formulated in~\cite{ZS1972} and~\cite{bk2014}, respectively.
Here we briefly present the essential steps of the IST for Eqs.~\eqref{e:MBE} 
with $q(t,z)\to q_\pm(z)$ as $t\to\pm\infty$ with $|q_\pm|= q_o>0$.
As with the NLS equation with NZBG~\cite{bk2014,ZS1973}, the eigenvalues of the scattering problem are branched
because they depend on $\gamma(\lambda) = (q_o^2 + \lambda^2)^{1/2}$.
One can deal with this issue by introducing a two-sheeted Riemann surface, obtained by gluing 
two sheets of the complex $\lambda$-plane in which $\gamma$ takes on the two signs of the square root,
with a branch cut on $i[-q_o,q_o]$.
On the first sheet, we take the principal branch of the square root, i.e., 
$\gamma(\lambda) = \sign(\lambda)\sqrt{q_o^2 + \lambda^2}$ for $\lambda\in\Real\cup i[-q_o,q_o]$,
so that the IST reduces to the case of ZBG as $q_o\to0$.
We also introduce, as in \cite{FT1987,bk2014},
the uniformization variable 
\vspace{-0.2ex}
\begin{equation}
\label{e:zeta}
\zeta = \lambda + \gamma\,,
\end{equation}
which is inverted by
\vspace{-0.2ex}
\begin{equation}
\label{e:lambdagamma}
\lambda = (\zeta - q_o^2/\zeta)/2\,,\quad
\gamma = (\zeta + q_o^2/\zeta)/2\,.
\vspace{-0.4ex}
\end{equation}
%
%
The formulation of the direct problem in the IST then proceeds 
in a manner which is essentially identical 
to that for the focusing NLS equation with NZBG \cite{bk2014},
by introducing the Jost solutions $\phi_\pm(t,z,\zeta)$ of the scattering problem,
the corresponding scattering data, which include reflection and transmission coefficients, 
as well as the discrete eigenvalues when applicable and the corresponding norming constants.
In particular, the scattering matrix $S(\zeta,z) = (s_{i,j})_{i,j=1,2}$ is defined by the relation $\phi_+(t,z,\zeta) = \phi_-(t,z,\zeta)S(\zeta,z)$ 
for $\zeta\in\Real$ and $|\zeta|=q_o$,
the reflection coefficient is $b(\zeta,z) = s_{2,1}/s_{1,1}$,
and the discrete eigenvalues $\zeta_1,\dots,\zeta_N$ are the zeros of $s_{1,1}$,
with associated norming constants $C_1,\dots,C_N$
\cite{bk2014}.

As in the MBEs with ZBG \cite{tmp1985}, the initial state of the medium is specified by assigning boundary conditions for the density matrix 
(consistently with causality)
as 
\begin{equation}
\rho_-(\lambda,z) = \lim_{t\to-\infty}\phi_\pm^{-1}(t,z,\lambda)\rho(t,z,\lambda)\,\phi_\pm(t,z,\lambda)\,.
\end{equation}

For general initial conditions, reconstructing the solution of the MBEs~\eqref{e:MBE} with NZBG
requires solving a Riemann-Hilbert problem~\cite{bk2014}.
Of course, as in the case of the MBEs with ZBG, the main difference with the NLS equation is in the spatial dependence of the scattering data.
A discussion of the general case is outside the scope of this work, and will be presented elsewhere \cite{IST4MBENZBC}.
In this work we limit ourselves to the case of pure soliton solutions, in which the formalism simplifies considerably.
We do so in the following sections.

\vglue\smallskipamount
\section{III.~ Soliton solutions and boundary conditions}
%
As usual, when the reflection coefficient is identically zero, the inverse problem can be solved in closed form,
and yields the $N$-soliton solutions of the system~\eqref{e:MBE} explicitly as
\begin{equation}
q(t,z) = \det M^\mathrm{aug}/\det M\,,
\label{e:Nsolitonsolution}
\end{equation}
where $M^\mathrm{aug}(t,z)$ and $M(t,z)$ 
(given explicitly in Appendix, together with the corresponding expression for $\rho(t,z,\lambda)$)
are completely determined in terms of the discrete eigenvalues of the scattering problem and the corresponding norming constants.
Equation~\eqref{e:Nsolitonsolution} is formally identical to the expression for the $N$-soliton solutions 
of the NLS equation with NZBG~\cite{bk2014}, 
but the resulting solutions are drastically different, as we discuss in section~V.
%
%
Similarly to the case of the MBEs with ZBG, pure soliton solutions can only exist when $\rho_-(\lambda,z)$ is diagonal, i.e.,
\vspace{-0.4ex}
\begin{equation}
\label{e:rho-}
\rho_-(\lambda,z) = \nu\, \sigma_3,
\end{equation}
where $\nu = \pm1$ correspond to atoms being initially in the excited state or the ground state, respectively.
With this choice,
\vspace*{-2ex}
\begin{gather}
D = \nu\,\lambda/\gamma + o(1)\,,
\quad
P = i\nu\,q_-/\gamma + o(1)\,,
\end{gather}
as $t\to-\infty$.
(Note that, unlike the case of ZBG, $P$ does not vanish in this limit.)\,\ 
Moreover,  
Eqs.~\eqref{e:MB2} and~\eqref{e:Lorentzian} imply that 
the NZBG of the optical field is independent of $z$ .
Thus, hereafter we take $q_-(z)=q_o>0$, 
without loss of generality.
Note that, unlike the case of ZBG, when $\rho_-$ is diagonal, the material polarization tends
to a constant, non-zero value as $t\to\pm\infty$.
Of course one can consider different choices for $\rho_-(\lambda,z)$, 
corresponding to different medium preparations.

We next briefly discuss this issue.
We show that some of these choices yield solutions in which the NZBG $q_-$ depends on $z$, 
and exhibit very different behavior than the ones presented below.
In particular, some of these alternative choices yield the solutions in Refs.~\cite{pb2001,xphc2016,hxpcd2016}.
However, as discussed below, 
the choice in \eqref{e:rho-} appears to be more natural from a physical point of view.

Some soliton solutions of the MBEs~\eqref{e:MBE} with NZBG had previously been obtained using direct methods~\cite{pb2001,xphc2016,hxpcd2016}.
As we discuss next, there are two main differences between those solutions and the ones presented in this work.

(i) Previous works only considered the sharp-line limit [i.e., the case $g(\lambda)$ is a Dirac delta],
whereas we consider a more general scenario as in \eqref{e:Lorentzian}, with the sharp-line limit being just a special case.
Thus, the formalism of the present work is more general.

(ii) All solutions presented in Refs.~\cite{pb2001,xphc2016,hxpcd2016} correspond to a NZBG for the optical field $q(t,z)$ 
that depends on $z$ [namely, $q_- = q_-(z)$], and to a more complicated background $\rho_-(\lambda,z)$
for the density matrix $\rho(\lambda,t,z)$.
Such choices yield to essential differences from a physical point of view for the resulting solutions.
Next we elaborate on this issue.
To do so, we need to first discuss how the choice of $\rho_-(\lambda,z)$ affects the solutions.

It is relatively straightforward to show that, with a general choice of $\rho_-(\lambda,z)$ and a general detuning function $g(\lambda)$,
the quantity $q_-(z)$ is fully determined from the limit of the MBEs~\eqref{e:MBE} as $t\to-\infty$ as follows:
\bse
\label{e:q-}
\begin{gather}
q_-(z) = q_o \e^{i W(z)}\,,\\
\noalign{\noindent where}
W(z) = \int_0^z w(z')\d z'\,,\\
\label{e:w}
w(z) = \int_\Real \rho_{-,1,1}(\lambda,z) g(\lambda) \sign(\lambda)\frac{\d\lambda}{\sqrt{\lambda^2 + q_o^2}}\,,
\end{gather}
\ese
with $\rho_{-,1,1}(\lambda, z)$ being the $(1,1)$-element of $\rho_-(\lambda, z)$, consistently with the notation used before.

Recall that $\lambda = 0$ is the normalized resonance frequency for the MBEs~\eqref{e:MBE},
and that the choice of $\rho_-(\lambda,z)$ corresponds to the initial preparation of the atoms.
The simplest possible choice is of course that in which $\rho_-(\lambda,z)$ is independent of $\lambda$.
Even if one considers situations in which $\rho_-(\lambda,z)$ depends on $z$, 
the most natural scenario is that in which $\rho_-(\lambda,z)$ is an even function of~$\lambda$,
unless special physical considerations dictate otherwise.
When $g(\lambda)$ is an even function of $\lambda$, it is easy to see from \eqref{e:w} that 
$w_-(z)=0$,
which in turn implies that $q_-(z)$ is also independent of~$z$.

Here we present solutions corresponding to physically relevant choices for $\rho_-(\lambda,z)$,
e.g., \eqref{e:rho-}.
One could consider alternative choices in which $\rho_{-,1,1}(\lambda,z)$ is an arbitrary function of $\lambda$.
Such choices can yield non-zero values of~$w_-(z)$ 
and therefore non-trivial dependence of $q_-(z)$ on $z$ via \eqref{e:q-}.
In turn, the different asymptotic behavior can result in very different solutions.
In Appendix we show solutions with an alternative choice of $\rho_-$,
which indeed exhibit quite different behavior, which is similar to some solutions recently presented in the literature.

As with the NLS equation~\cite{ZS1972,bk2014} and the MBEs~\eqref{e:MBE} with ZBG~\cite{akn1974},
one can show that all discrete eigenvalues are independent of $z$.
We consider $N$ distinct discrete eigenvalues $\lambda_j= \lambda(\zeta_j)$ in the upper-half plane with corresponding norming constants $C_j(z)$,  parameterized respectively by
\vspace*{-0.6ex}
\begin{equation}
\label{e:zetaC}
\zeta_j = q_o \, \eta_j \, \e^{i \alpha_j}\,,\quad
C_j(z) = \e^{\xi_j(z) + i\varphi_j(z)}\,,
\end{equation}
with $\eta_j >1$ and $0 < \alpha_j < \pi$,  
and with $\xi_j$ and $\varphi_j$ real.
Similarly to the NLS equation with NZBG, 
the symmetries of the scattering problem imply that,
in addition to any discrete eigenvalue $\zeta_j$ in the upper-half plane outsdide the circle of radius $q_o$ (i.e., with $|\zeta_j|>q_o$),  
an additional discrete eigenvalue $\zeta_{N+j} =  -q_o^2/\zeta_j^*$ inside the circle
is also present in the scattering problem, 
together with their complex complex conjugates $\zeta_j^*$ and $\zeta_{N+j}^*$ in the lower-half plane.
The additional norming constants are $C_{N+j}(z) = -(q_o/\zeta_j^*)^2 C_j^*(z)$, together with their symmetric counterparts.
Similarly to the case of ZBG, the evolution of the norming constants 
is determined by 
\vspace*{-0.4ex}
\begin{equation}
\label{e:dCdz}
\partial C_j / \partial z = -i\, \nu\, R(\zeta_j)\, C_j\,,
\end{equation}
but where now 
\vspace*{-0.2ex}
\begin{subequations}
\label{e:RTheta}
\begin{gather}
\label{e:Rzetaj}
R(\zeta_j) = g(\lambda_j)\big[\Theta(\lambda_j) - \gamma_j\,\Theta(i\epsilon)\big/(q_o^2 - \epsilon^2)^{1/2}\big]\,,\\
\hspace*{-0.4em}
\Theta(\lambda) = 2 \, \mathrm{arcsech}(-i \lambda/q_o)\,,
\label{e:Theta}
\end{gather}
\end{subequations}
with $\lambda_j = \lambda(\zeta_j)$ 
and $\gamma_j = \gamma(\zeta_j)$ 
evaluated using Eqs.~\eqref{e:lambdagamma}.
Equation~\eqref{e:dCdz} is solved immediately to give
\vspace*{-0.2ex}
\bse
\label{e:xivarphi}
\begin{gather}
\label{e:xi_j}
\xi_j(z) = \nu \, R_\mathrm{im}(\zeta_j) \, z + \xi_{j,0}\,,\\
\label{e:varphi_j}
\varphi_j(z) = -\nu \, R_\mathrm{re}(\zeta_j) \, z + \varphi_{j,0}\,,
\end{gather}
\ese
where the subscripts ``re'' and ``im'' denote the real and imaginary part, respectively,
$\xi_{j,0}\in\Real$ and $\varphi_{j,0}\in[-\pi,\pi)$.

\begin{figure}[t!]
\vglue\smallskipamount
    \centering
    \raisebox{2mm}{
    \includegraphics[scale=0.19]{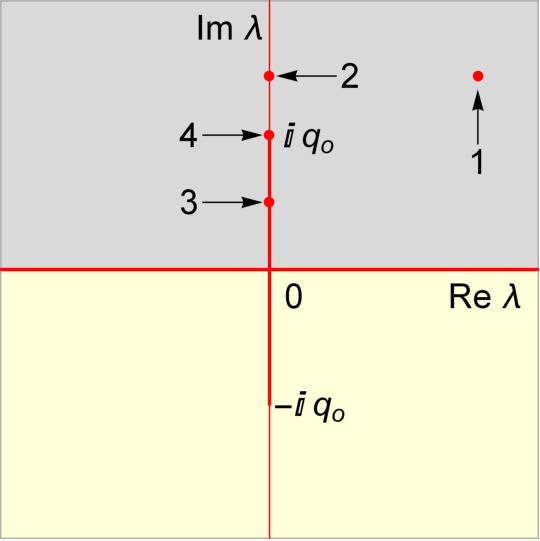}}~
    \includegraphics[scale=0.17,trim=0.1cm 0.1cm 0.1cm 0.1cm,clip]{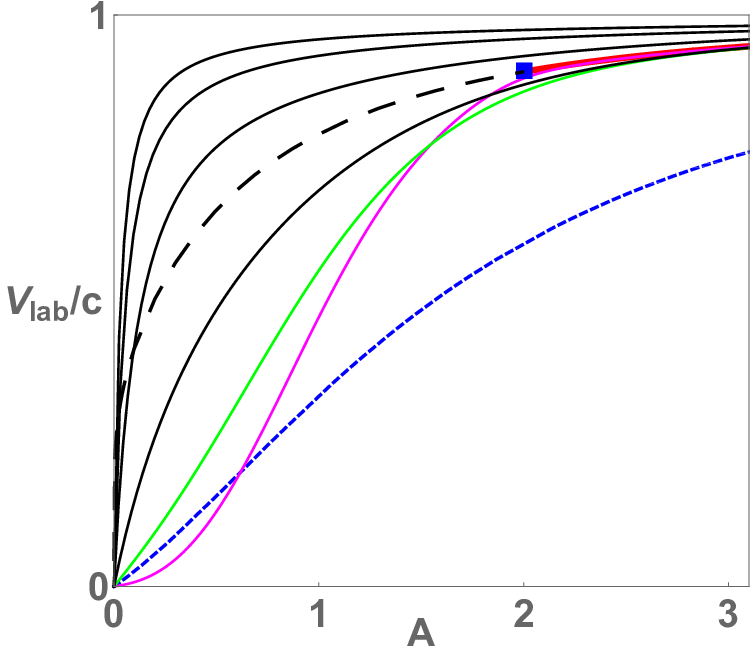}
    \caption{
        Left: the four types of discrete eigenvalues:
        I.~Oscillatory solitons;
        II.~Traveling-wave solitons;
        III.~Periodic solutions;
        IV.~Rational solutions.
        When $\Im\lambda_1>q_o$, type-I solutions limit to type~II as $\Re\lambda_1\to0$.
        When $\Im\lambda_1< q_o$ or $\Im\lambda_1 = q_o$ instead, they limit to type~III or type~IV, respectively.
        Right: soliton velocity in the lab frame 
        versus maximum soliton amplitude 
        for $\epsilon = 0.5$. 
        Red: $\Re\lambda_1 = 0$ and $\Im\lambda_1>1$.
        Blue square: $\lambda_1 = i$ (rational solutions).
        Dashed black: $\Re\lambda_1 = 0$ and $\Im\lambda_1<1$ (periodic solutions).
        Solid black (from bottom to top): $\Re\lambda_1 = 1,\dots,4$.
        Green: $\Re\lambda_1 = 0.5$.
        Magenta: $\Re\lambda_1 = 0.1$.
        For comparison, the dashed blue curve shows the soliton velocity in the case of ZBG when $\Re\lambda_1 = 0$.
        Here and in all figures below, $q_o=1$ and $\nu = -1$, corresponding to atoms in the ground state.
}
    \label{f:eigenvalue}
\vspace*{3ex}
    \centering
    \includegraphics[width=0.49\textwidth]{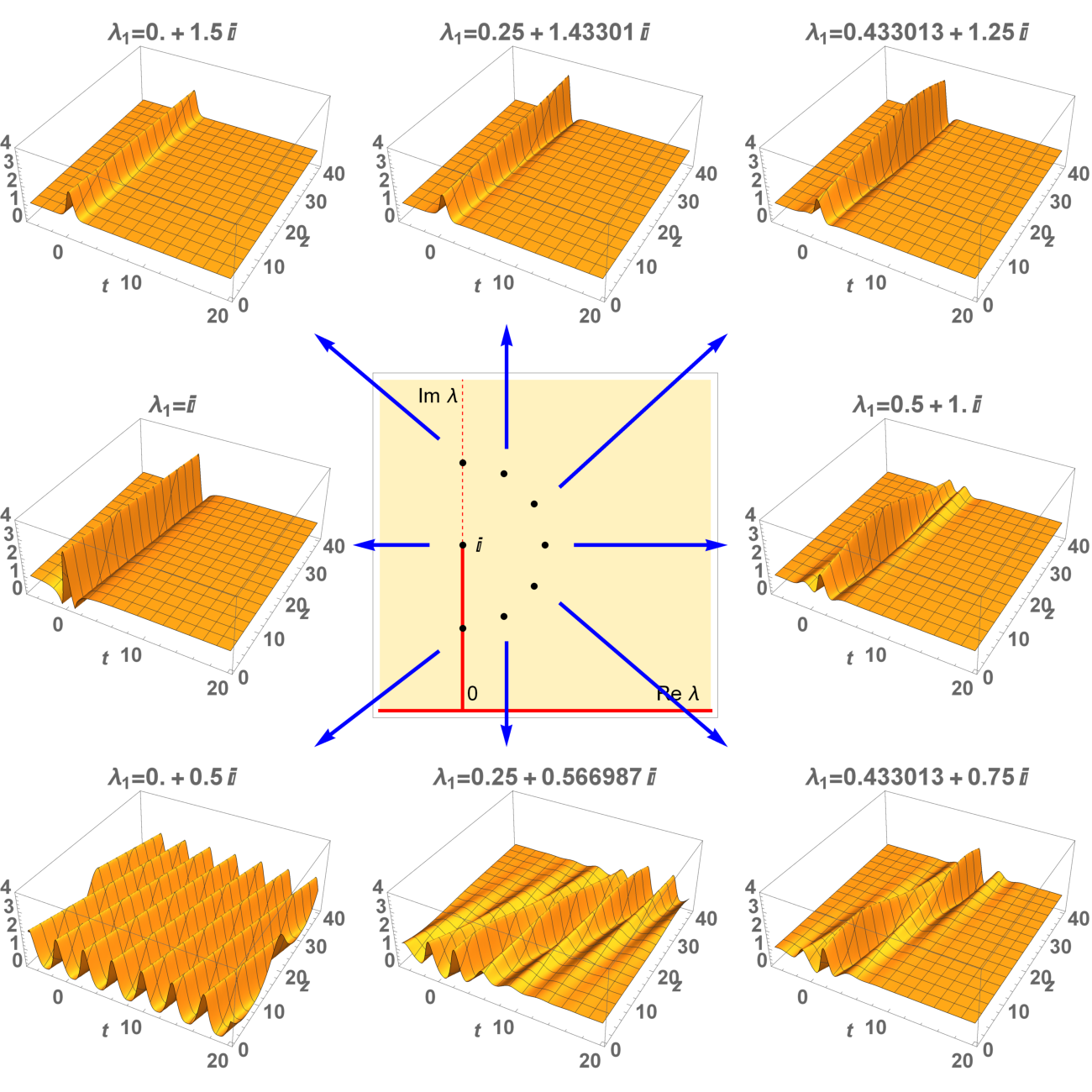}
    \caption{
Diagram illustrating the solution dependence on the location of the discrete eigenvalue $\lambda_1$.
The central plot shows eight discrete eigenvalues (black dots) in the spectral $\lambda$-plane.
Blue arrows indicate the correspondence between the location of these eigenvalues and the corresponding soliton solutions.
Left column from top to bottom: solutions of Type-II, IV and III, respectively.
All other five solutions are of type-I.
Note how, as the discrete eigenvalue evolves about the brunch point $\lambda = i$
(corresponding to figures from the upper left corner to the lower left corner, clockwise),
the single peak (Type-II) expands out (Type-I) and finally becomes a periodic solution (Type-III).
}
\vspace*{-4ex}
\label{f:summary}
\end{figure}

Below we use these expressions to discuss explicitly 
various soliton solutions of the MBEs~\eqref{e:MBE} with NZBG.
For simplicity we limit ourselves to the simplest case of one-soliton solutions ($N=1$),
but we will see that even in this case,
\eqref{e:Nsolitonsolution}
gives rise to four types of solutions,
identified by the location of the discrete eigenvalue $\lambda_1$, as shown in Fig.~\ref{f:eigenvalue}(left).

Figure~\ref{f:summary} presents a visual compendium of how the various solutions depend on the location of the discrete eigenvalue in the IST.
The corresponding solutions are described in detail in Section~IV.
The generic solution is obtained when the discrete eigenvalue is in general position in the spectral plane, 
and represents a breather-like structure comprising a hyperbolic envelope plus trigonometic oscillations.
This solution is described in Section~IV.A.
Distinguished limits of this general case are obtained when the discrete eigenvalue approaches the imaginary axis.
The corresponding solutions depend on whether the limiting point is above the branch cut or on it.
In the first limit (imaginary eigenvalue above the branch cut)
the internal oscillations disappear and one obtains traveling-wave solutions describing a solitary wave on top of the NZBG.
The corresponding solutions are described in Section~IV.B.
In the second limit (eigenvalue on the branch cut),
the width of the envelope tends to infinity, and one obtains periodic traveling-wave solutions.
These solutions are described in Section~IV.C.
Finally, a further distinguished limit is obtained when the discrete eigenvalue tends to the branch point $iq_o$.
In this case one obtains rational solutions of the MBE on NZBG.
These solutions are described in Section~IV.D.

\vglue\smallskipamount

\section{IV.~ One-soliton solutions}

For brevity, here we give $q(t,z)$.
Corresponding expressions for $\rho(t,z,\lambda)$ are given in Appendix.
Also,
for brevity we drop the subscript ``1'' from the  soliton parameters.

\subsection{IV.A~ Type~I: Oscillatory solitons}

We first discuss solutions corresponding to discrete eigenvalues in generic position. 
Equation~\eqref{e:Nsolitonsolution} 
with $N=1$ yields
\begin{equation}
\label{e:breather}
q_{\mathrm{i}}(t,z) = \e^{-2 i \alpha } q_o
\frac{\cosh (\chi-2i \alpha) + d_{+,2} G_s + i d_{-,2} G_c}{\cosh\chi - 2 G_s}\,,
\end{equation}
where $d_{\pm,j}= \eta^j \pm 1/\eta^j$ for $j=1,2\,$, 
\begin{gather*}
\kern-1em
\chi(t,z) = q_o d_{-,1} t\sin\alpha  + \ln \big[d_{+,1}/(2 q_o G_o \sin\alpha)\big] + \xi(z)\,,\\
G_s(t,z) =  \sin\alpha\,[\eta ^2 \sin (s-2 \alpha )+\sin s]/(G_od_{+,1})\,,\\
G_c(t,z) =  \sin\alpha\,[\eta ^2 \cos (s-2 \alpha )+\cos s]/(G_od_{+,1})\,,\\
G_o = |1 + \e^{2 i \alpha } \eta ^2|\,,\\
s(t,z) = -q_o d_{+,1} t\cos\alpha+\varphi(z)\,,
\end{gather*}
and 
where we used the identity
\vspace*{-0.6ex}
\begin{equation}
\cosh (a - ib) = \cosh a\cos b + i \sinh a \sin b\,.
\nonumber
\end{equation}
Two such solutions are displayed in Fig.~\ref{f:breather}.
The corresponding density matrices for all solutions discussed here are shown in Appendix.

The solution~\eqref{e:breather} describes a non-stationary oscillatory excitation traveling on top of the uniform background, 
with temporal frequency 
\begin{equation}
\nonumber
\omega = q_o|d_{-,1}\sin\alpha\,R_\re(\zeta_1)/R_\im(\zeta_1) - d_{+,1} \cos\alpha|/(2\pi)\,.
\end{equation}
This oscillatory behavior describes a cyclic transfer of energy between the light and the medium.
The solution is localized along the line $\chi(t,z) = 0$, 
corresponding to a velocity
\begin{equation}
V = -q_o \nu d_{-,1} \sin\alpha/R_\mathrm{im}(\zeta_1)\,,
\label{e:velocity}
\end{equation}
in the light-cone frame. 
Because of the breather-like nature of the solution~\eqref{e:breather},
$V$ is the group velocity of the structure.
The phase velocity (i.e., the velocity of each peak) is 
$V_p = -\nu q_od_{+,1}\cos\alpha/R_\re(\zeta_j)$.
%
The physical velocity $V_\mathrm{lab}$ of the soliton is then 
$V_\mathrm{lab} = V/(1+V/c)$.
Correspondingly, positive and negative velocities in the light-cone frame yield subluminal and superluminal motions, respectively.
Figure~\ref{f:eigenvalue}(right) shows $V_\mathrm{lab}$ as a function of the maximum soliton amplitude 
$A= \max|q(t,z)| - q_o$ from the NZBG.
Note that $V$ depends monotonically on~$A$.
That is, smaller solitons travel more slowly (as with ZBG \cite{mccallhahn2}),
and in particular $V\to0$ as $A\to0$.
Also, similarly to the case of ZBG, 
solitons are subluminal or superluminal
when atoms are initially in the ground state or the excited state, respectively.
Figure~\ref{f:amplitudevelocity} shows the soliton amplitude $A$ and velocity $V$ as individual functions of the 
discrete eigenvalue $\zeta_1$.

\begin{figure}[t!]
	\centering
	\includegraphics[scale=0.22,trim=0cm 0.25cm 0cm 0cm,clip]{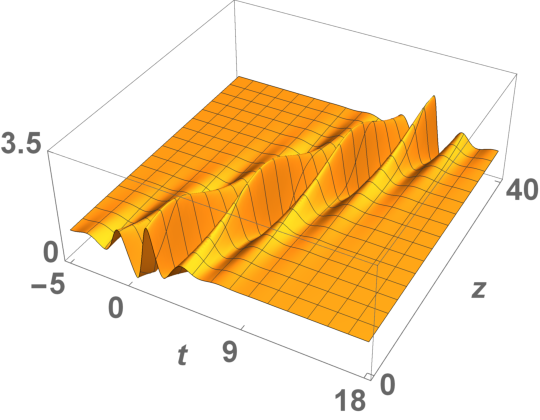}
	\includegraphics[scale=0.22,trim=0cm 0.25cm 0cm 0cm,clip]{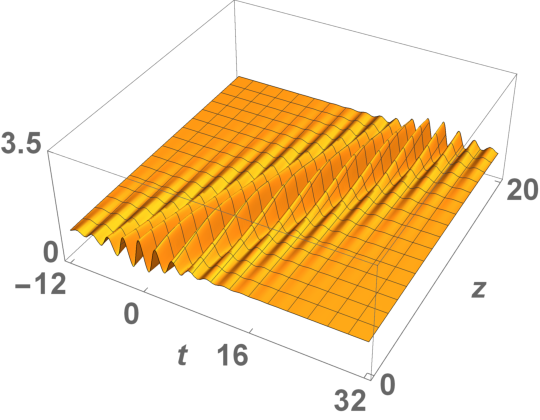}
	\vspace*{-1ex}
    \caption{
    Amplitude $|q(t,z)|$ of two type-I oscillatory soliton solutions given by \eqref{e:breather}
    with discrete eigenvalues $\zeta_1 = 2\,\e^{\pi i/6}$ (left),
    $\zeta_1 = 2\,\e^{\pi i/16}$ (right),
    detuning parameter $\epsilon = 2$ and norming constants $\xi_0 = \varphi_0 = 0$.}
	\label{f:breather}
\kern2\medskipamount
    \centering
    \includegraphics[scale=0.15]{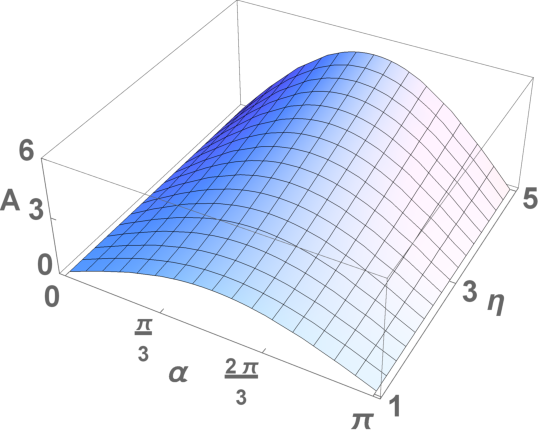}%
    \includegraphics[scale=0.15]{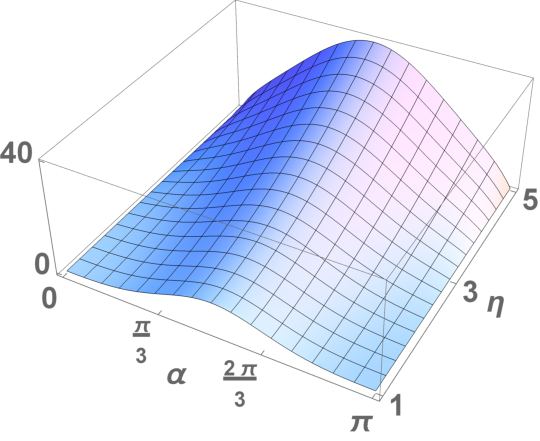}%
    \includegraphics[scale=0.15]{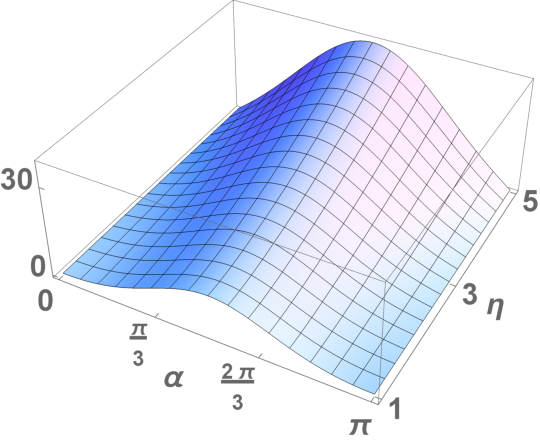}
    \caption{
        Left: maximum amplitude of type-I oscillatory solitons [\eqref{e:breather}] as a function of the eigenvalue parameters $\alpha$ and~$\eta$ [cf.\ \eqref{e:zetaC}].
        Center and right: velocity of oscillatory solitons described by \eqref{e:velocity}
        with $\epsilon = 0.5$ (center) and $\epsilon = 2$ (right).
        Other parameters are: $q_o = 1$, $\xi_0 = \varphi_0 = 0$ and $\nu = -1$.
        }    
    \label{f:amplitudevelocity}
\kern2\medskipamount
\centering
    \includegraphics[scale=0.22,trim=0cm 0.25cm 0cm 0cm,clip]{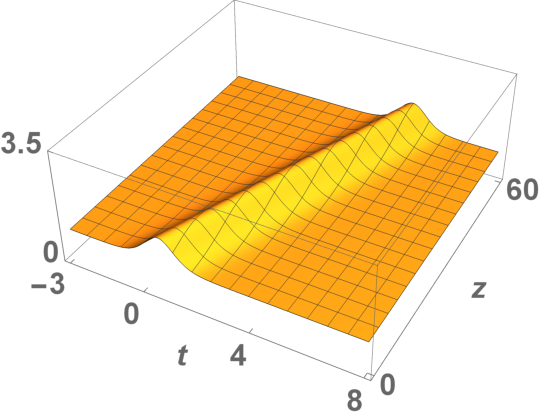}
    \includegraphics[scale=0.22,trim=0cm 0.25cm 0cm 0cm,clip]{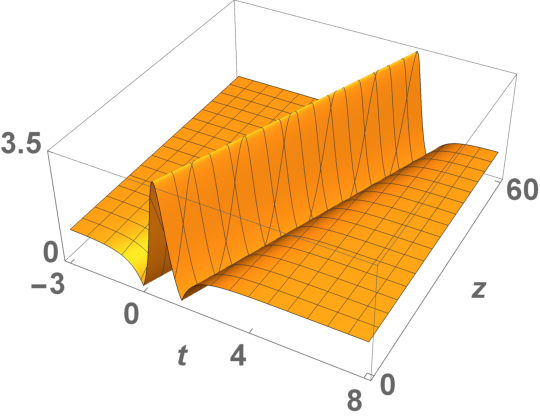}
    \vspace*{-1ex}
    \caption{
        Same as Fig.~\ref{f:breather} but for two type-II traveling-wave soliton solutions given by \eqref{e:static}, both
        with the same discrete eigenvalue $K = 2i$,
        illustrating how the norming constant affects the soliton amplitude and shape.
        Left: $\varphi_0 = 0$;
        Right: $\varphi_0 = -\pi/2$.
        All other parameters are the same as in~Fig.~\ref{f:breather}.
    }
    \label{f:static}
\vspace*{3ex}
	\centering
	\includegraphics[scale=0.22,trim=0cm 0.25cm 0cm 0cm,clip]{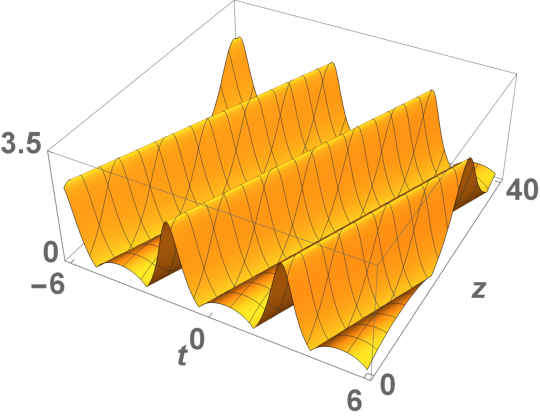}
	\includegraphics[scale=0.22,trim=0cm 0.25cm 0cm 0cm,clip]{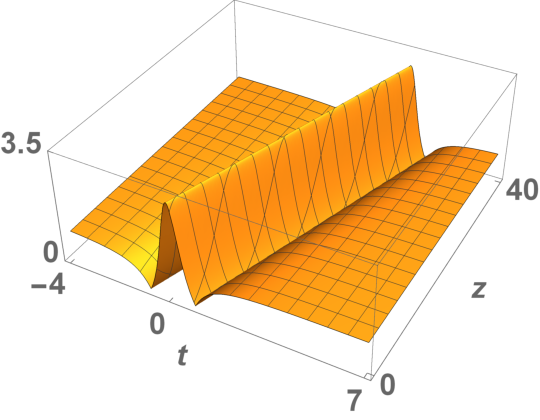}
	\vspace*{-1.5ex}
	\caption{
        Same as Fig.~\ref{f:breather} but for a type-III periodic solution given by \eqref{e:periodic} (left) with $\alpha = \pi/4$,
		and a type-IV rational solution in \eqref{e:rational} (right).
		All other parameters are the same as in Fig.~\ref{f:breather}.}
	\label{f:periodic}
	\vspace*{-2ex}
\end{figure}

One can show that 
in the sharp-line limit (i.e., as $\epsilon\to0$), $R(\zeta_1)\to0$,
and as a result the solution~\eqref{e:breather} becomes independent of $z$.
Hence, $V = \infty$, implying that the soliton travels with $V_\mathrm{lab}=c$, the speed of light in vacuum, in this limit.
However, soliton solutions for the MBEs with ZBG
have nontrivial $z$~dependence in the sharp-line limit~\cite{akn1974,bgk2003}.
Therefore, the two limits $\epsilon\to0$ and $q_o\to0$ do not commute.
Physically, this means that \textit{situations in which $\epsilon<q_o$ 
yield qualitatively different dynamics
from those in which $\epsilon > q_o$.}

One can also show that $\omega\to\infty$ as $\epsilon\to0$,
implying that the solitons in the sharp-line limit do not oscillate
(similarly to the solitons in the case of ZBG).
Hence, the internal structure that produces these oscillations is a new feature resulting from the combined presence of inhomogeneous broadening in systems with NZBG.

\vglue\smallskipamount
\subsection{IV.B~ Type II: Traveling-wave solitons on NZBG}

We now consider limiting cases of \eqref{e:breather}, which yield solution types II--IV in Fig.~\ref{f:eigenvalue}.
Consider first the special case of a purely imaginary discrete eigenvalue $\zeta_1 = i q_o\eta$, 
i.e., $\alpha = \pi/2$.
In this situation $R_\mathrm{re}(\zeta_1) = 0$, 
implying $\varphi(z) = \varphi_0$,
and the solution in \eqref{e:breather} reduces to
\begin{equation}
\label{e:static}
q_{\mathrm{ii}}(t,z) = q_o
	\frac{d_{+,1}\,\cosh\chi +  d_{+,2} \sin \varphi_0 +  i d_{-,2}\cos \varphi_0}
	{d_{+,1}\,\cosh\chi + 2\,\sin \varphi_0}\,,
\end{equation}
with 
\begin{equation}
\chi(t,z) = q_od_{-,1}t + \ln[d_{+,1}/(2q_od_{-,1}\eta)] + \xi(z) 
\end{equation}
and $\xi(z)$ still given by \eqref{e:xi_j}.
Two such solutions are shown in Fig.~\ref{f:static}.

Unlike the type~I solutions, solitons given by \eqref{e:static} have an invariant temporal profile, 
like the solitons of the MBEs with ZBG. 
(This is because as $\alpha\to\pi/2$ the temporal dependence of $s(t,z)$ vanishes.)
\,\ 
Unlike the case of ZBG,
the maximum amplitude $A$ from the background depends not only on the discrete eigenvalue, 
but also on the norming constant, namely
$A = q_o\big\{\big[1+ d_{-,1}d_{-,2}/(d_{+,1} + 2\sin\varphi_0)\big]^{1/2} - 1\big\}$. 
The maximum, $A_\mathrm{max} = q_o d_{+,1}$, is achieved for $\varphi_0 = -\pi/2$.

\vglue\smallskipamount
\subsection{IV.C~ Type~III: Periodic solutions}

Special solutions are also obtained 
in the limit when the discrete eigenvalue approaches the branch cut. 
More precisely, 
as $\eta\to1$ with $\alpha\ne\pi/2$, 
which corresponds to the discrete eigenvalue $\lambda(\zeta_1)$ approaching the segment $[0,iq_o]$,
\eqref{e:breather} yields
\vspace*{-1ex}
\begin{equation}
\label{e:periodic}
q_{\mathrm{iii}}(t,z) \! = \! \e^{-2i\alpha} \! q_o
	\big[\cosh(\chi + 2i\alpha) - K \big]\big/(\cosh\!\chi + K)\,,
\end{equation}
where 
$\chi = \ln(q_o|\sin2\alpha|) - \xi_0$ is now constant, and
\begin{gather*}
s(t,z) = 2q_ot \cos \alpha + \nu R_\epsilon\,z - \varphi_0\,,\\
K(t,z) = \sign(\pi/2-\alpha)\sin(s + \alpha)\sin\alpha\,,\\
R_\epsilon = g(iq_o\sin\alpha)
\bigg[2\,\mathrm{arcsech}(\sin\alpha) - \frac{q_o \cos\alpha\,\Theta(i\epsilon)}{\sqrt{\smash{q_o}^2-\epsilon^2}}\bigg]\,.
\vspace*{0.5ex}
\end{gather*}
As~$\eta\to1$, $\chi$ becomes independent of $t$ and $z$.
Thus the only $(t,z)$-dependence of the solution arises from $s(t,z)$, which only appears in trigonometric functions. 
As a result,
\eqref{e:periodic} is periodic with respect to both the spatial and temporal variables, with temporal frequency 
$\omega = q_o\cos\alpha/\pi$,
and spatial period
$L = 2\pi/R_\epsilon$.
One such solution~\eqref{e:periodic} 
is shown in Fig.~\ref{f:periodic}(left).

The envelope of a type~I solution stretches out to infinity as $\eta\to1$.
The limit, given by \eqref{e:periodic},
describes a periodic transfer of energy between the light and the medium.
The solution~\eqref{e:periodic} does not tend to a constant value for $q(t,z)$.
This is similar to what happens for the Akhmediev breathers in the NLS equation~\cite{bk2014}.
Nonetheless, \eqref{e:Nsolitonsolution} also holds for $\eta=1$, in which case it yields
\eqref{e:periodic}.
In this limit, the group velocity loses its meaning,
and the phase velocity is simply
$V_p = -2q_o \nu \cos\alpha/R_\epsilon\,$.

Once more, $R_\epsilon\to0$ as $\epsilon\to0$, and the resulting solution does not depend on $z$.
Moreover, the solution vanishes in the ZBG limit, i.e., as $q_o\to0$.
In other words, the periodic solution is a novel feature arising from the presence of the NZBG.

\vglue\smallskipamount
\subsection{IV.D~ Type~IV: Rational solutions}

Further exact solutions are obtained 
when the discrete eigenvalue approaches the branch point, i.e., 
$\zeta_1\to iq_o$.
This corresponds to the limit $\eta\to1$ of the solution~\eqref{e:static} or equivalently to the limit
$\alpha\to\pi/2$ of the solution~\eqref{e:periodic},
in which case $\omega\to0$ (i.e., the resulting solution becomes localized in $t$).
By appropriately rescaling of the norming constant,
which ensures that the peak of the soliton is kept at a fixed location in the limiting process,
this limit yields the following rational solutions of the MBEs~\eqref{e:MBE}:
\vspace*{-0.4ex}
\begin{equation}
\label{e:rational}
q_{\mathrm{iv}}(t,z) = q_o \big(\chi^2 + \varphi_0^2 + 4i \varphi_0 - 3\big)\!\big/\!
	\big(\chi^2 + \varphi_0^2 + 1\big),
\end{equation}
where
\vspace*{-1ex}
\begin{equation}
\chi(t,z) = 2q_o t + \nu \, g(i q_o) \, \tilde R_\epsilon \,z + \xi_0
\end{equation}
and
$\tilde R_\epsilon = 2 - q_o\Theta(i\epsilon)\big/\sqrt{q_o^2-\epsilon^2}$.
Equation~\eqref{e:rational} describes a traveling-wave solution of the MBEs on NZBG, with
maximum amplitude 
$A = q_o\big[(\varphi_0^4 + 10 \varphi_0^2 + 9)^{1/2}/(\varphi_0^2 + 1) - 1\big]$
and velocity 
$V = -2\nu q_o/[\,g(i q_o)\tilde R_\epsilon]$.
An example of this solution is shown in Fig.~\ref{f:periodic}(right).
In the sharp-line limit, $g(i q_o)\to0$,
implying that the solution becomes independent of $z$ and travels with the speed of light in vacuum.
It is also easy to show that this rational solution vanishes in the ZBG limit.

\vglue\smallskipamount
\section{V.~ Discussion}
%

From a spectral point of view, solutions of types~I--IV are the MBE analogue 
of the Tajiri-Watanabe~\cite{tajiriwatanabe},
Kuznetsov-Ma~\cite{Kuznetsov,Ma} 
Akhmediev~\cite{akhmediev} 
and Peregrine~\cite{peregrine} 
solitons of the NLS equation, respectively.
But the behavior of the solutions in the two models is very different.
(Type~II solitons are traveling-wave solutions,
whereas Kuznetsov-Ma solitons are periodic in space and localized in time.
Similarly, 
type~III solutions are periodic in both space and time,
whereas Akhmediev breathers are periodic in time and localized in space.
Finally, type~IV solitons are traveling-wave solutions, whereas Peregrine solitons are localized in both space and time.)

Compared to the MBE with ZBG,
the addition of a NZBG drastically affects the behavior of the solutions,
which acquire an extra degree of freedom and in general are not traveling-wave solutions anymore,
but rather breather-like in nature, with a structure characterized by a modulated amplitude with 
distinct phase velocity and group velocity.
We have also shown that special limits of this general solution arise when the discrete eigenvalue approaches the imaginary axis,
in which case the structure either limits to a periodic solution 
(when the modulus of the discrete eigenvalue is less than that of the NZBG)
or reduces back to a traveling wave (in the opposite case).

From a practical point of view, the results of this work mean that one could approach each of these solutions
by tuning appropriate parameters in the experimental set-up.
We also reiterate that it is crucial to consider the case of inhomogeneous broadening as opposed to the sharp-line limit,
because in the latter case one could easily end-up with physically unrealizable solutions.
as discussed in section~III.

We also note that, 
even though for simplicity we only discussed one-soliton solutions, the formalism described in section~II
is quite general, and allows one to generate $N$-soliton solutions for arbitrary $N$, 
including solutions with combinations of the various kinds of eigenvalues.

\section{Acknowledgments}

This work was partially supported by the National Science Foundation under grant numbers DMS-1615524 and
DMS-1615859.

\addcontentsline{toc}{section}{Appendix}

\setcounter{section}1
\setcounter{subsection}0
\setcounter{equation}0
\def\thesection{\Alph{section}}
\def\theequation{\Alph{section}.\arabic{equation}}

\section{Appendix}

Here we provide further details on the general $N$-soliton solutions of the MBEs~\eqref{e:MBE},
as well as formulae and plots of the density matrix corresponding to the soliton solutions presented in the main text.
Additionally, we also provide solutions that corresponding to different choices of the boundary conditions for the density matrix.

\subsection{$N$-Soliton solution formulae and density matrices}
%

We use the same parameterization for the $N$ discrete eigenvalues $\zeta_1,\dots,\zeta_N$ and corresponding norming constants as in the main text. 
[Recall that the norming constants $C_1,\dots,C_N$ are given explicitly by Eqs.~\eqref{e:xivarphi} in the main text.]
Similarly to the nonlinear Schr\"odinger (NLS) equation with non-zero background (NZBG), 
the symmetries of the scattering problem imply that,
in addition to any discrete eigenvalue $\zeta_j$ in the upper-half plane outsdide the circle of radius $q_o$ (i.e., with $|\zeta_j|>q_o$),  
an additional discrete eigenvalue $\zeta_{N+j} =  -q_o^2/\zeta_j^*$ inside the circle
is also present in the scattering problem, 
together with their complex complex conjugates $\zeta_j^*$ and $\zeta_{N+j}^*$ in the lower-half plane.
where asterisk denotes complex conjugation.
The additional norming constants are $C_{N+j}(z) = -(q_o/\zeta_j^*)^2 C_j^*(z)$, together with their symmetric counterparts.

To write the $N$-soliton solution in compact form, it is convenient to define the scalar, vector and matrix quantities
\vspace*{-0.2ex}
\begin{gather*}
A = (A_{n,l})_{2N\times2N}\,,\quad 
\textbf{B} = (B_n)_{2N\times1}\,,\\
\textbf{C} = (C_n)_{2N\times1}\,,\quad
\textbf{D} = (D_n)_{1\times 2N}\,,\\
A_{n,l} = \sum_{j=1}^{2N} c_j(\zeta_n^*,z)c_l^*(\zeta_j^*,z)\,,
\end{gather*}
\eject
\vglue-2ex
\begin{gather*}
B_n = 1 + i q_o\sum_{j = 1}^{2N} c_j(\zeta_n^*,z)/\zeta_j\,,\\
C_n = iq_o /\zeta_n^* + \sum_{j=1}^{2N}c_j(\zeta_n^*,z)\,,\quad
D_n = -C_n^*(z)\,\e^{2i\gamma(\zeta_n^*)t}\,,\\
c_j(\zeta,z) = C_j(z)\,\e^{-2i\gamma(\zeta_j)t}/(\zeta-\zeta_j)\,,\quad
n,l = 1,\dots,2N\,.
\vspace*{-4ex}
\end{gather*}
Then the optical field $q(z,t)$ is given by Eq.~\eqref{e:Nsolitonsolution} in the main text with
\vspace*{-1ex}
\begin{equation}
\nonumber
M = I + A\,,\qquad
M^\mathrm{aug} = \begin{pmatrix}
	q_o & i \, \textbf{D} \\ \textbf{B} & M
\end{pmatrix}\,,
\vspace*{-1ex}
\end{equation}
where $I$ is the $2N\times2N$ identity matrix.
The corresponding density matrix is reconstructed using the relation
\begin{equation}
\label{e:rho}
\rho(t,z,\zeta) = \nu \, \mu(t,z,\zeta)\,\sigma_3\,\mu^{-1}(t,z,\zeta)\,,
\end{equation}
where $\nu = \pm1$ indicates the initial state of atoms,
$\sigma_3 = \mathrm{diag}(1,-1)$ is the third Pauli matrix, 
and the modified matrix eigenfunction $\mu(t,z,\zeta) = \phi(t,z,\zeta)\,\e^{-i\gamma t\sigma_3}$ is
given by
\begin{multline}
\mu(t,z,\zeta) = \begin{pmatrix} 1 & i q_o/\zeta \\ i q_o/\zeta & 1\end{pmatrix}
\\
+ \sum_{n=1}^{2N}\frac{C_n\,\e^{-2i\gamma(\zeta_n)t}}{\zeta-\zeta_n}\begin{pmatrix}\tilde{X}_n &0\\ \tilde{Y}_n &0\end{pmatrix}
\\
- \sum_{n=1}^{2N}\frac{C_n^*\,\e^{2i\gamma(\zeta_n^*)t}}{\zeta-\zeta_n^*}\begin{pmatrix}0 &X_n\\ 0 &Y_n\end{pmatrix}\,,
\end{multline}
where $X_n$, $Y_n$, $\tilde X_n$, $\tilde Y_n$ for $n = 1,\dots,2N$ 
are given by the solution of the following linear system:
\vspace{-0.5ex}
\begin{gather*}
M \,\textbf{X} = \textbf{B}\,,\qquad
M \,\textbf{Y} = \textbf{C}\,,\\
\begin{pmatrix}\tilde{X}_n\\\tilde{Y}_n\end{pmatrix} = 
\begin{pmatrix}iq_o/\zeta_n\\1\end{pmatrix} - \sum_{j=1}^{2N} c_j^*(\zeta_n^*,z)
\begin{pmatrix}X_j\\Y_j\end{pmatrix}\,. 
\end{gather*}
Below we give the first two entries of $\mu(t,z,\zeta)$ for all four types of solutions discussed in the main text,
[omitting the corresponding expressions for $\rho(t,z,\lambda)$ for brevity].
Note that for $\zeta\in\Real$, 
$\mu(t,z,\zeta)$ satisfies the symmetries $\mu_{2,1}(t,z,\zeta) = -\mu_{1,2}^*(t,z,\zeta)$ and $\mu_{2,2}(t,z,\zeta) = \mu_{1,1}^*(t,z,\zeta)$,
so the other two entries can be easily obtained.
Moreover, recall Eq.~\eqref{e:Qrho}, 
we will only plot $D(t,z,\lambda)$ and $|P(t,z,\lambda)|$ in all figures.
Importantly, notice that $\gamma(\lambda)$ is discontinuous at $\lambda = 0$, and as a result, so is the 
density matrix $\rho(t,z,\lambda)$. 
In all figures we show the entries of $\rho(t,z,\lambda)$ in the limit $\lambda\to0^+$.

\subsection{Type 1. Oscillatory solitons}

\begin{figure}[b!]
    \centering
    \includegraphics[scale=0.22]{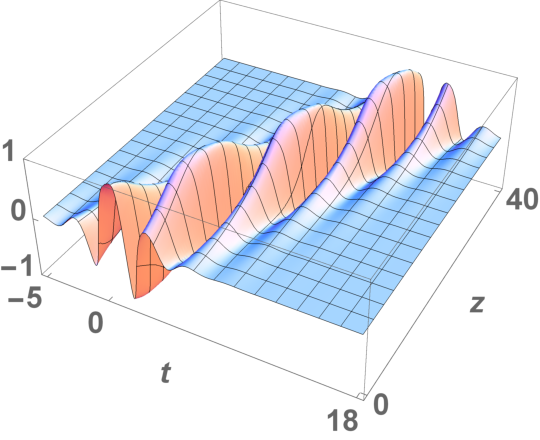}
    \includegraphics[scale=0.22]{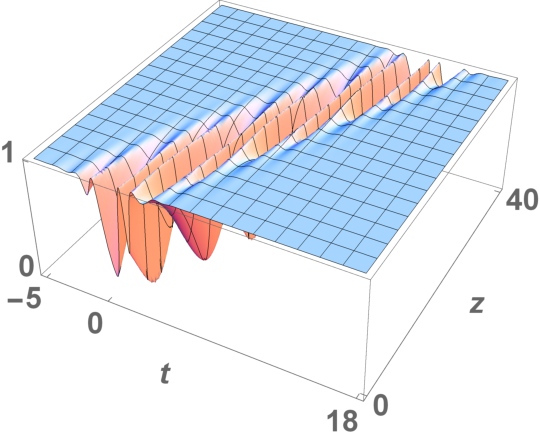}
    \includegraphics[scale=0.22]{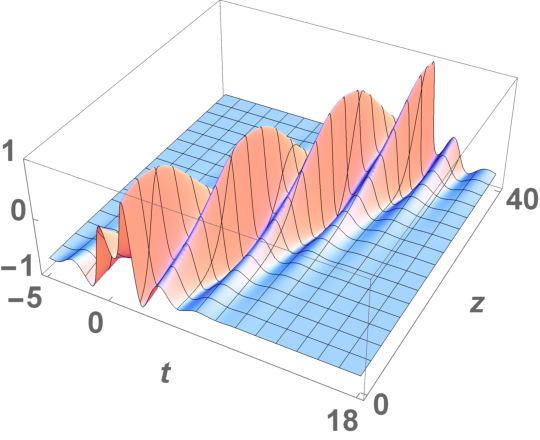}
    \includegraphics[scale=0.22]{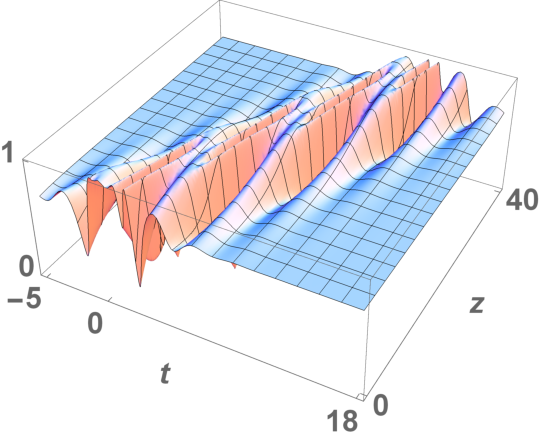}\\
    \includegraphics[scale=0.22]{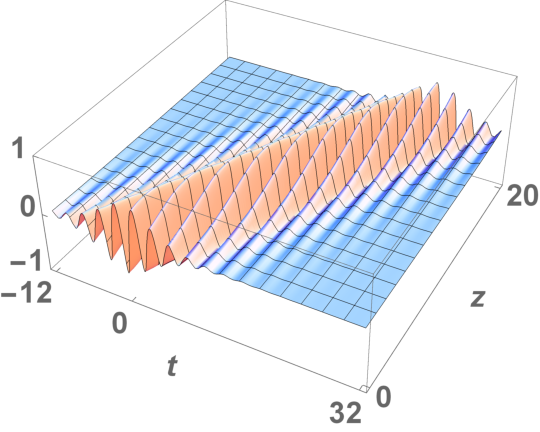}
    \includegraphics[scale=0.22]{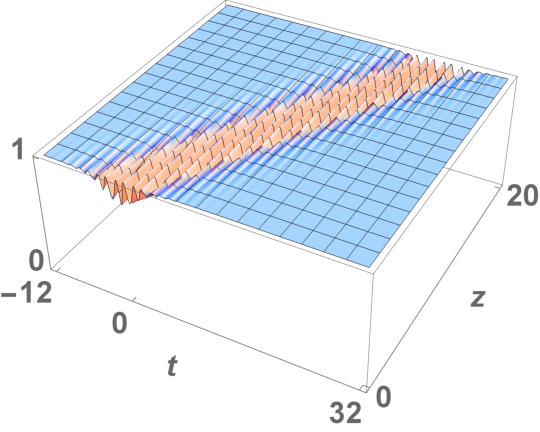}
    \includegraphics[scale=0.22]{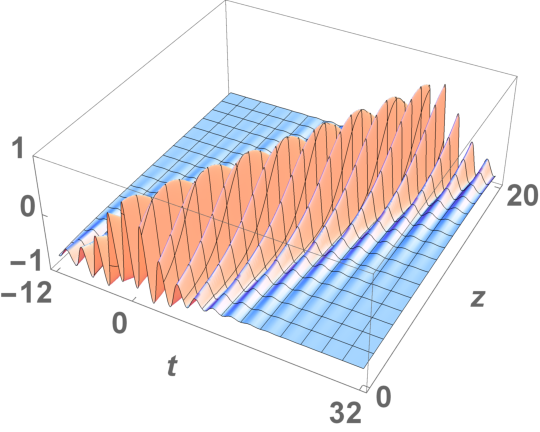}
    \includegraphics[scale=0.22]{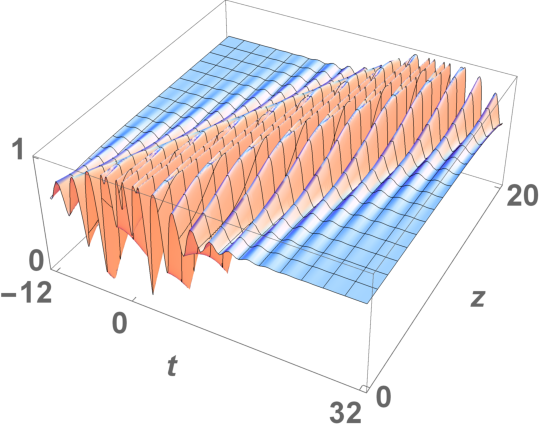}
    \caption{
        Entries of the density matrix $\rho(t,z,\lambda)$ for the type-I oscillatory soliton solution in Fig.~\ref{f:breather}(left) and (right) in the main text
        (top row and bottom row here, respectively).
        First column: $D(t,z,0^+)$.
        Second column: $|P(t,z,0^+)|$.
        Third column: $D(t,z,1)$.
        Last column: $|P(t,z,1)|$.
    }
    \label{f:breather_density}
\end{figure}

Recall the discrete eigenvalue is $\zeta_1 = \eta q_o \e^{i\alpha}$.
Introducing the shorthand notation $\hat{\zeta} = -q_o^2/\zeta$,
the matrix $\mu$ is
\begin{widetext}
\vspace*{-4ex}
\begin{subequations}
    \label{e:long}
    \begin{multline}
    \mu_{1,1} = \big[q_o^2 \cosh (\chi-2i\alpha) + d_{-,1} \zeta  q_o \cosh(\chi - i\alpha) - \zeta ^2 \cosh\chi\big]/(X Y)\\
    +i \sin\alpha\big(F_o \e^{-i s}|\zeta -\zeta_1|^2 - \e^{i s} F_o^* |\zeta -\hat\zeta_1| ^2\big)\big/(F d_{+,1} X Y) \,,
    \end{multline}
    \vspace{-6ex}
    \begin{multline}
    \mu_{1,2} = - i \e^{-2i\alpha} q_o\big[\zeta \cosh (\chi - 2i \alpha) - d_{-,1}  q_o \cosh(\chi - i\alpha) + \hat{\zeta} \cosh\chi\big]\big/
    (X Y^*)\\
    + \e^{-2i\alpha}  q_o\sin\alpha\big(\e^{-i s} \eta^4 F_o |\zeta - \hat\zeta_1| ^2-\e^{i s} F_o^* |\zeta -\zeta_1|^2\big)\big/
    (F d_{+,1} \zeta  \eta^2 X Y^*)\,.
    \end{multline}
\end{subequations}
\vspace*{-2ex}
\end{widetext}
where 
\begin{gather*}
F_o = 1 + \e^{2 i \alpha } \eta^2,\\
    X = \cosh\chi - 2 \kappa_s\,,\\
    Y = \e^{2 i \alpha } q_o^2 + \e^{i \alpha } d_{-,1} \zeta  q_o-\zeta ^2\,.
\end{gather*}
Two density matrices corresponding to the same choice of parameters that yield 
the soliton solutions in Fig.~\ref{f:breather} are shown in Fig.~\ref{f:breather_density}. 

\subsection{Type 2. Traveling wave solitons}
\begin{figure}[t!]
    \centering
    \includegraphics[scale=0.22]{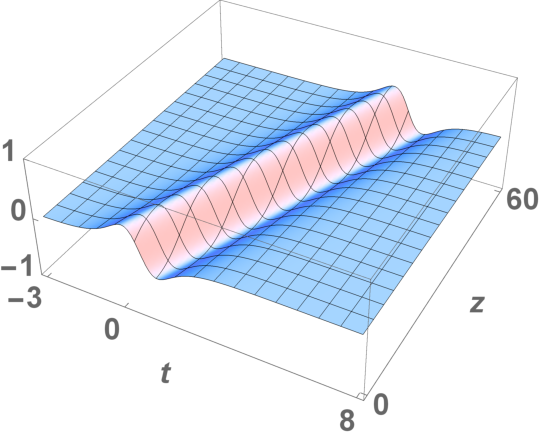}
    \includegraphics[scale=0.22]{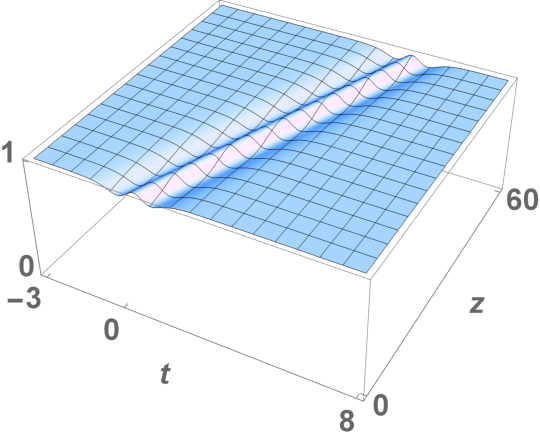}
    \includegraphics[scale=0.22]{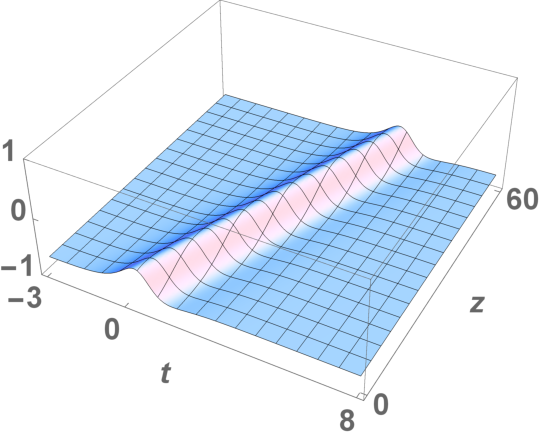}
    \includegraphics[scale=0.22]{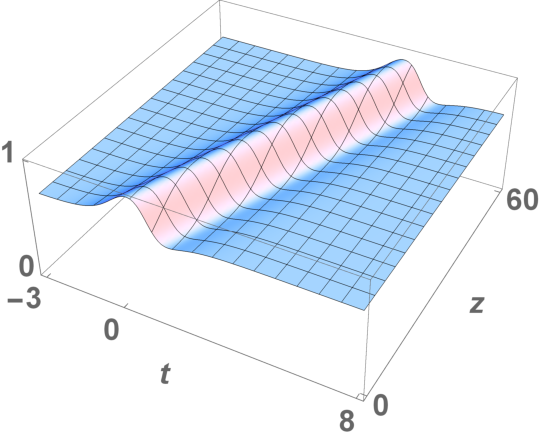}\\
    \includegraphics[scale=0.22]{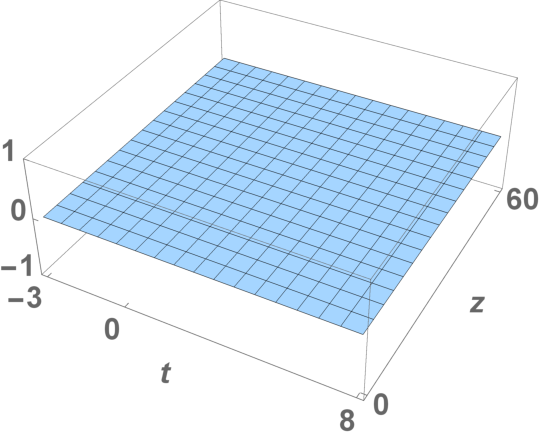}
    \includegraphics[scale=0.22]{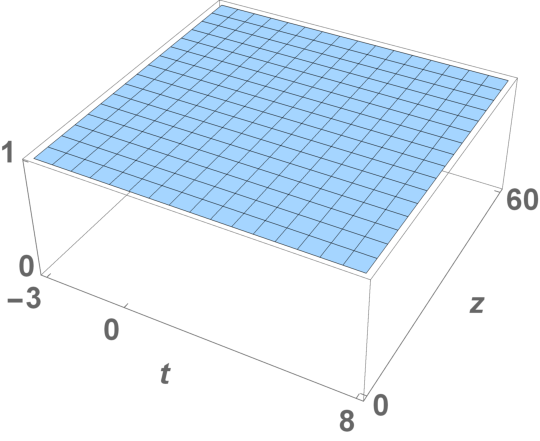}
    \includegraphics[scale=0.22]{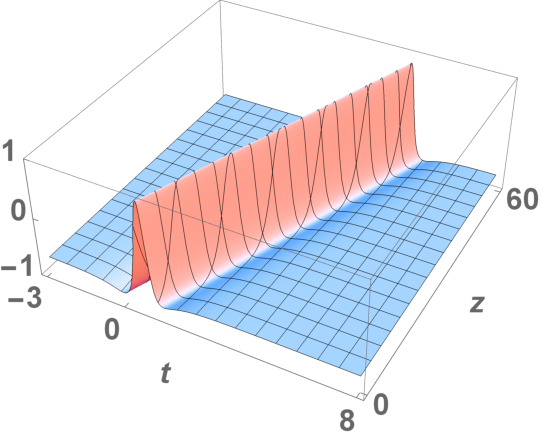}
    \includegraphics[scale=0.22]{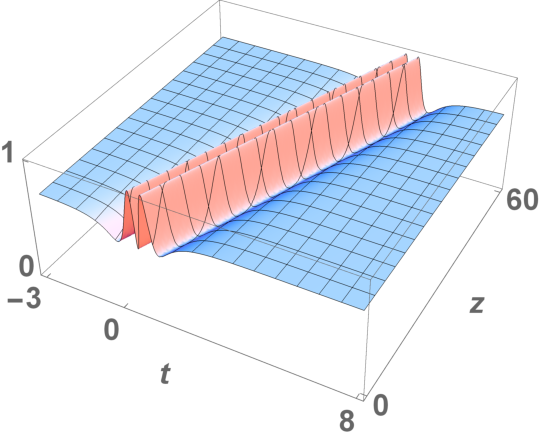}
    \caption{
        Same as Fig.~\ref{f:breather_density},
        but for the two type-II traveling wave soliton solutions in Fig.~\ref{f:static}(left) and (right) in the main text
        (top row and bottom row here, respectively).
    }
    \label{f:static_density}
\end{figure}

The discrete eigenvalue in this case is $\zeta_1 = i\eta q_o$.
The soliton solution for the optical field is given by Eq.~\eqref{e:static}.
The corresponding density matrix is obtained by Eq.~\eqref{e:rho}, with $\mu(t,z,\zeta)$ given by
\begin{subequations}
\begin{gather}
\mu_{1,1} = 
	(- d_{-,2} \zeta  q_o \sinh\chi + id_{-,1} q_o^2 Y + 2 i \gamma  \zeta  X)/(\zeta X Z)\,,\\
\mu_{1,2} = q_o (-i d_{-,2} q_o \sinh\chi + d_{-,1} \zeta Y + 2 \gamma  X)/(\zeta  X Z^* )\,,
\end{gather}
\ese
where
\bse
\begin{gather}
X = d_{+,1}\cosh\chi + 2\sin \varphi_0\,,\\
Y = d_{-,1} \sin \varphi_0 + id_{+,1} \cos \varphi_0\,,\\
Z = d_{-,1} q_o + 2 i \gamma\,.
\end{gather}
\end{subequations}
Two such density matrices are shown in Fig.~\ref{f:static_density} here,
corresponding to the two soliton solutions in Fig.~\ref{f:static} in the main text. 

\subsection{Type 3. Periodic solutions}
The discrete eigenvalue in this case is $\zeta_1 = q_o\e^{i\alpha}$.
The matrix $\mu(t,z,\zeta)$ is 
\bse
\begin{multline}
\mu_{1,1} = \big[\zeta^2 X - q_o^2 \cosh(\chi + 2 i\alpha) + q_o^2K +i \zeta q_o \tilde{K}\big]
\\
\big/\big[(\zeta ^2-\zeta_1^2)X \big]\,,
\end{multline}
\vspace{-2ex}
\begin{multline}
\mu_{1,2} =  i q_o \big[q_o^2 X - \zeta^2 \cosh (\chi + 2i\alpha) + \zeta ^2K
- i \zeta q_o \tilde{K}\big]
\\
\big/\big[\zeta (q_o^2 - \e^{2 i \alpha } \zeta ^2)X\big]\,,
\end{multline}
where
\begin{equation}
\tilde{K} = \sin (2 \alpha ) \cos(s+\alpha )\,,\qquad
X = \cosh\chi + K\,,
\end{equation}
\ese
where $\chi$, $s$ and $K$ are given in the main text.
One such solution is shown in Fig.~\ref{f:periodic_density}(left), 
corresponding to the solution in Fig.~\ref{f:periodic}(left) in the main text. 

\subsection{Type 4. Rational solutions}
\begin{figure}[b!]
	\centering
	\includegraphics[scale=0.22]{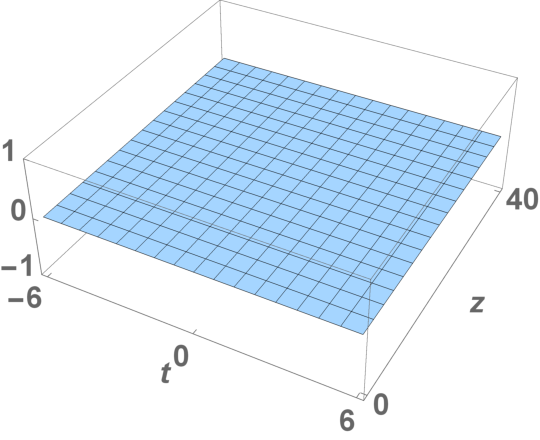}
	\includegraphics[scale=0.22]{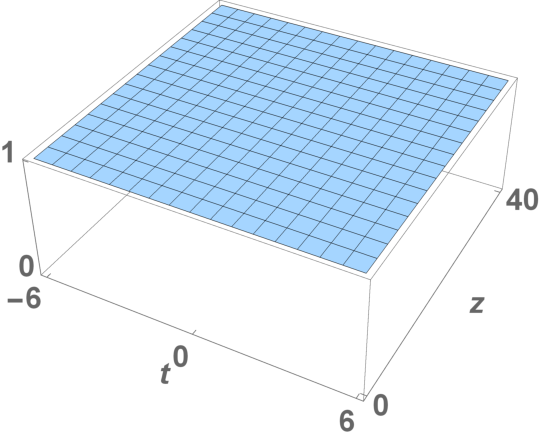}
	\includegraphics[scale=0.22]{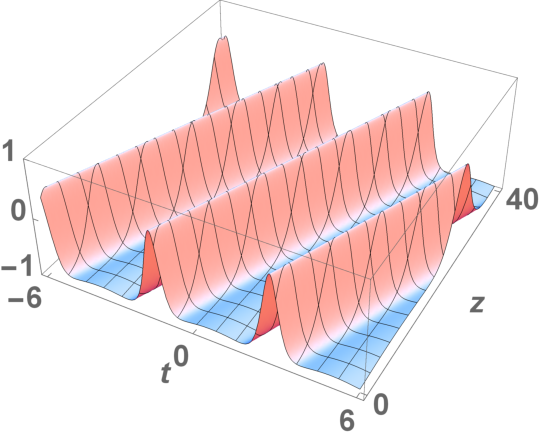}
	\includegraphics[scale=0.22]{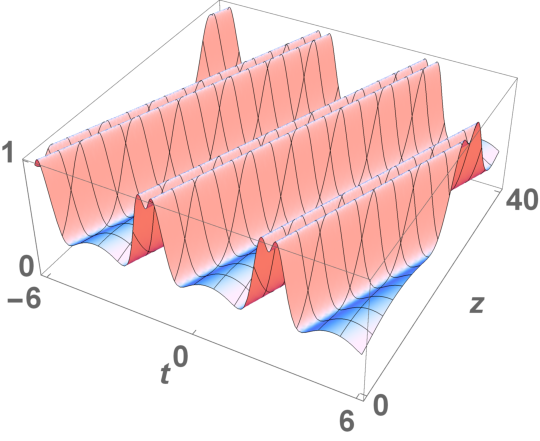}\\
	\includegraphics[scale=0.22]{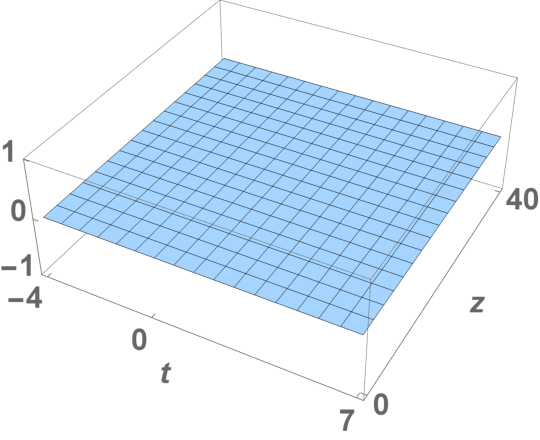}
	\includegraphics[scale=0.22]{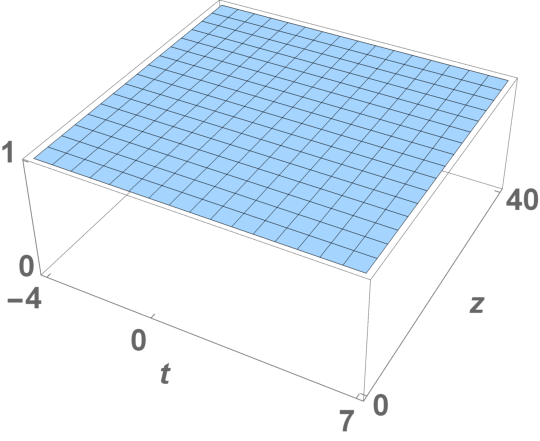}
	\includegraphics[scale=0.22]{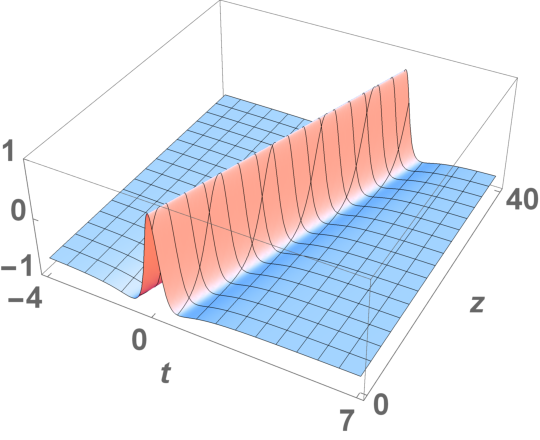}
	\includegraphics[scale=0.22]{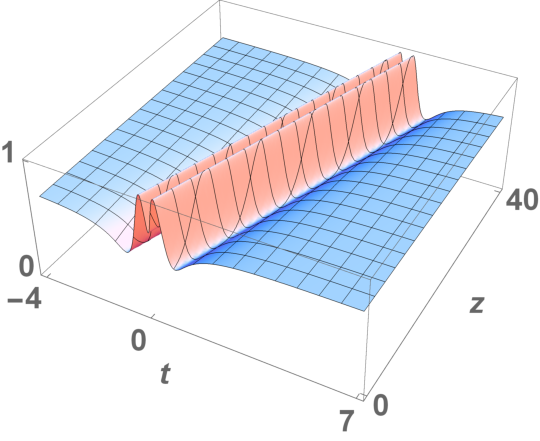}
\caption{
    Same as Fig.~\ref{f:breather_density}, 
		but for a type-III periodic solution (top row) and a type-IV rational solution (bottom row), 
		corresponding to the solutions in Fig.~\ref{f:periodic}(left and right) in the main text,
        respectively.
    }
\vspace*{-2ex}
	\label{f:periodic_density}
\end{figure}

The discrete eigenvalue in this case is $\zeta_1 = iq_o$.  
The matrix $\mu(t,z,\zeta)$ is now 
\bse
\begin{multline}
\mu(t,z,\zeta) = 
\big[
(Y\gamma\zeta-2q_o^2) I + 2iq_o(q_o \varphi_0 + \chi\zeta) \sigma_3
\\
+ i(Y \gamma - 2 \zeta)\sigma_3 Q_- +2(q_o \chi-\zeta \varphi_0) Q_-
\big]
/(\gamma  \zeta  Y)\,,
\end{multline}
where
\begin{equation}
Q_- = \begin{pmatrix} 0 & q_o \\ -q_o & 0 \end{pmatrix}\,,\qquad
Y = \chi^2 + \varphi_0^2+1\,,
\end{equation}
\ese
One such solution is shown in Fig.~\ref{f:periodic_density}(right), 
corresponding to Fig.~\ref{f:periodic}(right) in the main text.

\subsection{Boundary conditions for the density matrix and alternative solutions}

\begin{figure}[t!]
\kern\medskipamount
\centerline{\includegraphics[scale=0.22]{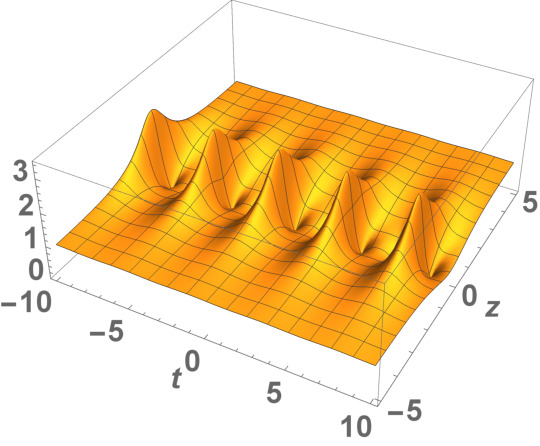}\includegraphics[scale=0.22]{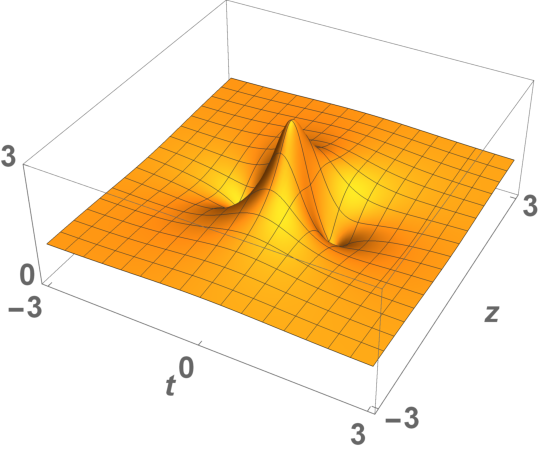}}
\caption{
        Amplitude $|q(t,z)|$ of two soliton solutions with $\nu = -1$ and the alternative BC~\eqref{e:rho-_alternative} in the sharp-line limit.
        Left: a Type-III periodic solution.
        Right: a Type-IV rational solution from Eq.~\eqref{e:qrational_alternative}.
    }
    \label{f:rational_alternative}
\end{figure}

Recall the discussion in the main text on different choices of the boundary conditions.
We hereby provide examples of soliton solutions with an alternative (less physically important) BC.
We consider the mathematically interesting solutions with the following choice of BC:
\begin{equation}
\label{e:rho-_alternative}
\rho_-(\lambda,z) = \sign(\lambda)\,\nu\,\sigma_3\,,\qquad \nu = \pm1\,.
\end{equation}
Such a BC would seem to be rather unphysical, because it corresponds to a situation in which the material preparation depends
on the detuning parameter
(with different signs according to the atoms' relative velocities),
Nonetheless, with this choice,
the evolution of the norming constant with respect to~$z$ [as determined by Eq.~\eqref{e:dCdz}] 
is very different from Eqs.~\eqref{e:RTheta} and~\eqref{e:xivarphi},
which are obtained with the more physical choice of Eq.~\eqref{e:rho-}.
Correspondingly, all four types of soliton solutions change dramatically.
For instance, in the sharp-line limit, the rational solution becomes
\begin{equation}
\label{e:qrational_alternative}
q(t,z) =  q_o \e^{i\nu z/q_o}\frac{4q_o^4 t^2 + 4 z^2 - 8i\nu q_o z - 3q_o^2}
{4q_o^4 t^2 + 4 z^2 + q_o^2}\,.
\end{equation}
The behavior of the solution in Eq.~\eqref{e:qrational_alternative} 
(which is qualitatively similar to that of the solutions in Refs.~\cite{xphc2016,hxpcd2016})
is quite different from Eq.~\eqref{e:rational} 
[cf.\ Fig.~\ref{f:periodic}(right)],
and is surprisingly similar to that of the Peregrine soliton of the NLS equation~\cite{peregrine}.
An example of a solution of type~III and a solution of type~IV generated by the BC in Eq.~\eqref{e:rho-_alternative}
is shown in Fig.~\ref{f:rational_alternative}.

\def\reftitle#1,{\relax}



\end{document}